# Crystal structure and anisotropic magnetic properties of new ferromagnetic Kondo lattice compound Ce(Cu,Al,Si)$_2$

A. Maurya[1], A. Thamizhavel[1], S. K. Dhar[1], A. Provino[2,3], M. Pani[2,3], G. A. Costa[2,3]

[1]Department of Condensed Matter Physics & Materials Science, Tata Institute of Fundamental Research, Homi Bhabha Road, Mumbai 400 005, India

[2]Department of Chemistry, University of Genova, Via Dodecaneso 31, 16146 Genova, Italy

[3]Institute SPIN-CNR, Corso Perrone 24, 16152 Genova, Italy

**Abstract**

Single crystals of the new compound CeCu$_{0.18}$Al$_{0.24}$Si$_{1.58}$ have been grown by high-temperature solution growth method using a eutectic Al-Si mixture as flux. This compound is derived from the binary CeSi$_2$ (tetragonal α-ThSi$_2$-type, Pearson symbol $tI$12, space group $I4_1/amd$) obtained by partial substitution of Si by Cu and Al atoms but showing full occupation of the Si crystal site (8$e$). While CeSi$_2$ is a well-known valence-fluctuating paramagnetic compound, the CeCu$_{0.18}$Al$_{0.24}$Si$_{1.58}$ phase orders ferromagnetically at $T_C$ = 9.3 K. At low temperatures the easy-axis of magnetization is along the $a$-axis, which re-orients itself along the $c$-axis above 30 K. The presence of hysteresis in the magnetization curve, negative temperature coefficient of resistivity at high temperatures, reduced jump in the heat capacity and a relatively lower entropy released up to the ordering temperature, and enhanced Sommerfeld coefficient (≈100 mJ/mol K$^2$) show that CeCu$_{0.18}$Al$_{0.24}$Si$_{1.58}$ is a Kondo lattice ferromagnetic, moderate heavy fermion compound. Analysis of the high temperature heat capacity data in the paramagnetic region lets us infer that the crystal electric field split doublet levels are located at 178 and 357 K, respectively, and Kondo temperature (8.4 K) is of the order of $T_C$ in CeCu$_{0.18}$Al$_{0.24}$Si$_{1.58}$.

**Keywords**: Ce(Cu,Al,Si)$_2$, Single crystal, Ferromagnetic Kondo lattice, Heavy fermion, Crystal electric field.

## 1. Introduction

Heavy fermion Kondo lattice compounds are one of the most intriguing subjects in the correlated electron physics. They have been extensively studied to probe the interplay of competing RKKY (Ruderman-Kittel-Kasuya-Yosida) and single-ion Kondo exchange interactions, non-Fermi liquid behavior, quantum criticality and unconventional superconductivity. The majority of heavy fermion Kondo lattice compounds that order magnetically are antiferromagnetic, while those that order ferromagnetically are far less in number [1]. Antiferromagnetic Kondo lattices respond similarly when they are tuned by external pressure, magnetic field or by alloying, their ferromagnetic counterparts apparently do not follow a generic behavior. It is, therefore, of interest to discover and explore new ferromagnetic Kondo lattices.

Material exploration by crystal growth has been proven to be useful to discover new materials. By using the high temperature solution growth technique and the eutectic mixture of Al-Si (87.4 : 12.6 at.%) as flux, we were recently successful in growing single crystals of the quaternary $RTAl_4Si_2$ for R = Ce, Pr and Eu (R = rare earth) and T = Ir and Rh [2-4]. By adopting the same protocol, we attempted to grow single crystals for T = Cu. Though the procedure resulted in the formation of single crystals, a detailed characterization showed that the crystals were pertaining to a pseudo-binary $Ce(Cu_xAl_ySi_{1-x-y})_2$ phase, a derivative of the tetragonal $CeSi_2$ compound, in which Si is partly substituted by both Cu and Al atoms. The $CeSi_{2-x}$ phase exhibits a tetragonal α-$ThSi_2$-type crystal structure [Pearson code *tI*12; space group $I4_1/amd$ (No. 141)] and a homogeneity region of $1:78 \leq x \leq 2:0$; according to literature it melts congruently for x = 1:86 at a temperature of 1620 °C [5,6].

The stoichiometry of our single crystal is $CeCu_{0.18}Al_{0.24}Si_{1.58}$. In this report we describe the crystal structure and magnetic behavior of this new compound by means of single crystal and powder x-ray diffraction (XRD), magnetization, electrical transport and heat capacity data. In $CeSi_{2-x}$ compounds, a transition from paramagnetic spin/valence fluctuations to ferromagnetic dense Kondo lattice occurs with decreasing Si concentration. [7, 8, 9]. Our work shows that $CeCu_{0.18}Al_{0.24}Si_{1.58}$ is a Kondo lattice ferromagnet with $T_C$ = 9.3 K.

## 2. Experimental techniques

The single crystal of $CeCu_{0.18}Al_{0.24}Si_{1.58}$ was grown by using flux technique and the eutectic $Al_{87.4}Si_{12.6}$ mixture as excess solvent which has a melting point of 575°C. A mixture of Ce, Cu, Al and Si, in the atomic composition ratio 1:1:30:6.8 at.%, was placed in an alumina crucible. The crucible was further sealed inside a quartz ampoule after it was evacuated to a pressure of ≈$10^{-6}$ torr; the temperature of the ampoule was gradually ramped up to 1100°C in 24 hours and maintained at that value for 24 hours to ensure the homogenization of mixture. The molten charge was then cooled down to 675°C at a rate of 2°C/h; at this temperature the excess flux was centrifuged out. Shining single crystals were present in the residual solid mass; well defined Laue spots confirmed the good quality of crystals. Several single crystals were checked for size and morphology by using a scanning electron microscope (SEM), equipped with an electron dispersive x-ray analysis detector (EDAX) for semi-quantitative analysis. The crystals were fixed on a substrate by a carbon tape. A counting time of 60 seconds was used; the estimated accuracy was ±0.5 at.% for Ce and within ±0.7 at.% for each of the other elements. The phase purity of the crystals was further checked by x-ray powder diffraction using a PANalytical x-ray diffractometer with monochromated Cu K$_\alpha$ radiation. Single-crystal intensity data for structural refinement were collected at 293 K on a Bruker-Nonius MACH3 diffractometer, using graphite-monochromated Mo K$_\alpha$ radiation. Crystals to be measured were oriented by means of Laue diffraction using the Huber Laue diffractometer. A polycrystalline non-magnetic reference sample, with composition LaCu$_{0.18}$Al$_{0.24}$Si$_{1.58}$, was prepared by melting the constituents in an arc furnace under an inert pure-Ar atmosphere. The phase purity of the alloy was confirmed by x-ray powder diffraction. Susceptibility and isothermal magnetization measurements were performed in the Quantum Design (QD) superconducting quantum interference device (SQUID) magnetometer and QD vibration sample magnetometer (VSM). The heat capacity, electrical resistivity and transverse magnetoresistance data, in the range 1.8-300K, were collected in the QD physical properties measurement system (PPMS). Heat capacity down to 50 mK was measured in the dilution refrigerator attachment of QD PPMS.

## 3. Results
### 3.1. Crystal Structure

Multiple SEM-EDAX analyses gave values of the Ce content centered around 33.5 at.%, with an average composition of the crystals resulting to be about 1:0.18:0.24:1.58 for the elemental ratios of Ce:Cu:Al:Si; these results led to an overall stoichiometry of Ce(Cu,Al,Si)$_2$, with formula CeCu$_{0.18}$Al$_{0.24}$Si$_{1.58}$. Laue patterns revealed that the crystals have a tetragonal symmetry; the powder x-ray diffraction pattern could be indexed by a tetragonal unit cell with lattice parameters $a = 4.211(4)$ Å and $c = 14.117(9)$ Å. These values are comparable with those of the well-known compound CeSi$_{2-x}$ ($0 \leq x \leq 0.4$) ($a = 4.184$ Å and $c = 13.856$ Å for CeSi$_2$) [7], which crystallizes in the tetragonal $\alpha$-ThSi$_2$-type structure. It may be noted that the bigger Si atoms ($V = 20.02$ Å$^3$) are substituted by Cu and Al atoms which have a smaller atomic volume (11.81 Å$^3$ and 16.60 Å$^3$ for Cu and Al, respectively) [10]. Small single crystals, apt for single-crystal work, were picked out after crushing a larger crystal in silicone oil and checked by Laue method; one of the best then selected for data collection and structural investigation. Crystal data and structure refinement details for CeCu$_{0.18}$Al$_{0.24}$Si$_{1.58}$ are given in Table 1; the standardized fractional atomic coordinates (with displacement parameters) and bond distances are collected in Tables 2 and 3, respectively. Ce and M atoms (M= Cu/Al/Si) occupy the Wyckoff sites *4a* and *8e*, respectively, with full site occupation. There are no Ce-Ce bond distances and/or even relatively short Ce-Ce lengths; only Ce-M and M-M bonds are present (see Table 3).

As reported in Ref. [7], the crystal structure of ThSi$_2$ is known to be found only for silicides and germanides; the sublattice structure constructed by Si (or Ge in CeGe$_2$), which may be looked upon as a 3-dimensional graphite structure, is important in forming this structure type. Isostructural pseudo-binary alloys of composition CeSi$_{2-x}$T$_x$ (with T = Al, Ga, Ge, Cu), as well as quaternary single crystals with composition Ce(Ag, Al, Si)$_2$ have also been reported in the literature [11-13].

Figure 1 shows projections of the unit cell of the CeCu$_{0.18}$Al$_{0.24}$Si$_{1.58}$ along the [100] (Fig. 1a), [010] (Fig. 1b) and [001] (Fig. 1c) directions, where the polyhedra around the Ce atoms are highlighted.

### 3.2. Magnetization

Figures 2a and 2b show the magnetization, *M*(*T*), below 20 K, measured in the zero-field-cooled (ZFC) and field-cooled (FC) modes and in applied magnetic fields of 0.05 T and 0.1 T for field parallel to the *a* and *c*-axis, respectively. The FC data strongly suggest a transition to a ferromagnetic state at T ≈10 K, with a strong anisotropy well evident between the two crystallographic directions [100] and [001]. The ZFC data show the presence of a peak, which is typical of ferromagnets when the magnetization curve is recorded in an applied field smaller than the coercive field, whose scale is primarily set by the magnetocrystalline anisotropy. In this regard it may be noted that when the field is parallel to *a*-axis the peak is well evident for μ$_0$H = 0.05 T, while it has nearly vanished for μ$_0$H = 0.1 T (there is only a downturn at *T*< 3K) (Figure 2a); when the field is parallel to the *c*-axis, the peak is present both at 0.05 and 0.1T (Figure 2b).

Figure 2c shows the inverse magnetic susceptibility, $\chi^{-1}$, between 1.8 and 300 K, measured in an applied field of 0.1 T for field parallel to the [100] and [001] directions, respectively. The inverse susceptibility is strongly anisotropic in the entire range of temperature, as already observed from the data of *M*(T) at low temperatures (Figures 2a and 2b). Above 100 K, the following Curie-Weiss expression:

$$\chi = \frac{N\mu_B^2 \mu_{eff}^2}{3k_B(T-\theta_P)}$$

where the various symbols have their usual meaning, provides a good fit to the data giving $\mu_{eff}$ = 2.18 $\mu_B$/Ce and 2.53 $\mu_B$/Ce and $\theta_P$ = −42.3 K and −29.5 K for applied field oriented parallel to the [100] and [001] directions, respectively. Along the tetragonal *c*-axis, the $\mu_{eff}$ value is nearly the same as that expected for the trivalent free ion $Ce^{3+}$ (2.54 $\mu_B$), but the value is smaller along the *a*-axis. The paramagnetic Curie temperature $\theta_P$ is negative along both the directions, suggesting *prima-facie* that the exchange interaction between the Ce ions, mediated via the RKKY exchange interaction, is antiferromagnetic. The ferromagnetic like response that sets in below ≈10 K is in contradiction to the expectation of an antiferromagnetic transition indicated by the negative value of $\theta_P$. It may be noted that Kondo interaction can also give a negative contribution to $\theta_P$; we believe that the negative value of $\theta_P$ in this compound arises due to the dominant contributions made by the crystal electric fields and single ion Kondo exchange interaction, even though the compound orders ferromagnetically. Interestingly, the susceptibility data suggest a possible change of the easy axis of magnetization from the tetragonal *c*-axis [001] to the *a-b* plane (001) at around ≈40 K where the two plots cross each other. This is amply confirmed by the *M*(T) data where the magnetic response for *H* oriented parallel to the [100] direction (*i.e.* along the *a*-axis) is larger by more than one order of magnitude (Figure 2a) compared to that for *H* oriented parallel to the [001] direction (*i.e.* along the *c*-axis) (Figure 2b). Further confirmation is provided by the data shown in Figure 2d, which plots the isothermal magnetization at *T* = 2 K along the two main directions for fields up to 16 T. The magnetization rises quickly at low fields along the *a*-axis while it increases gradually along the *c*-axis. The nearly saturated magnetization of ≈ 0.66 $\mu_B$/Ce along [100] is lower than the value of 2.14 $\mu_B$ for trivalent free $Ce^{3+}$ ion. Part of the reduction in the saturation magnetization can be attributed to the crystal electric field effect. However, Kondo effect, which arises due to an appreciable hybridization between the Ce-4*f* orbital with the conduction electrons, may also result in such a reduction. The possibility of Kondo interaction in the present compound is suggested by our resistivity data which exhibit a negative temperature coefficient of resistivity above $T_C$ and up to 300 K (*see below*). The inset in Figure 2d shows the expanded hysteresis loop between 1 T and −1 T. A coercive field of nearly 0.09 T and 0.6 T for *H* ∥ [100] and *H* ∥ [001], respectively, is inferred from the width of the loops, which provides an explanation for the effect of the applied field on the peaks seen in the ZFC *M*(T) data in Figures 2a and 2b.

### 3.3. Electrical resistivity

The electrical resistivity as a function of temperature, $\rho(T)$, and transverse magnetoresistance, *MR*(*H*) in $CeCu_{0.18}Al_{0.24}Si_{1.58}$ are shown in Figure 3, for the configuration *J* ∥ [100] and *H* ∥ [010]. The resistivity of the La sample decreases monotonically with temperature (not shown) as the sample is cooled below 300 K, which is typical of a paramagnetic metal. On the other hand, a negative temperature coefficient is observed in CeCu$_{0.18}$Al$_{0.24}$Si$_{1.58}$ below 300 K down to $T_C$ ~ 9.5 K, where a sharp peak in the resistivity is a signature of the magnetic transition inferred above from the magnetization data (Figure 3a main panel). $\rho$(T) of both alloys is observed to drift to a higher value with temperature cycling; possibly due to the gradual development of thermally assisted cracks in the sample. However, this drift is smaller in polycrystalline LaCu$_{0.18}$Al$_{0.24}$Si$_{1.58}$ but the magnitude of $\rho(T)$ in it is higher than that of single crystalline CeCu$_{0.18}$Al$_{0.24}$Si$_{1.58}$. This does not allow to estimate the 4$f$- electron contribution to the electrical resistivity, $\rho_{4f}$ in the Ce-analog, which is typically derived by subtracting the resistivity of the La-analog from the corresponding Ce-compound. CeCu$_{0.18}$Al$_{0.24}$Si$_{1.58}$ exhibits a negative slope in its resistivity behavior from 300 K right down to T$_C$, which is a characteristic feature in Kondo lattice compounds.

The $\rho(T)$ at selected fields has been plotted in the inset of Figure 3a for $J \parallel$ [100] and $H \parallel$ [010]. The peak in the zero field is increasingly broadened as the field is increased. Below a temperature $T$, which increases in magnitude with increasing field, the resistivity is lower than its zero field value. The magnetoresistance, $MR$, defined as $MR(H) = (\rho(H)-\rho(0)) \times 100/\rho(0)$ for the transverse configuration ($J \parallel$ [100] and $H \parallel$ [010]) is calculated from the $\rho(H)$ data and shown in Figure 3b. At 2K, the $MR$ of CeCu$_{0.18}$Al$_{0.24}$Si$_{1.58}$ is negative, and attains a value of -7.6% at 14 T, arising from the combined effect of the field on the Kondo many body state and residual ferromagnetic spin fluctuations. Note that the $MR$ of La analog (not shown) is an increasing function with $\mu_0 H$ with a value of 0.3 % at 14 T field. The negative curvature in $MR$ of CeCu$_{0.18}$Al$_{0.24}$Si$_{1.58}$ is enhanced as $T_C$ is approached, leading to a $MR$ of -14.4% at 14 T and 10 K. At higher temperatures, $MR$ shows a positive curvature at low fields (see 12, 15, and 25 K traces), and becomes positive at 50 K. The minimum in $MR$ in the vicinity of $T_C$ may be explained by the combined effect of Kondo effect and suppression of ferromagnetic fluctuations near $T_C$ by the magnetic field.

### 3.4. Heat Capacity

Heat capacity as a function of temperature, $C_p(T)$, was measured from 0.05 K to 300 K for CeCu$_{0.18}$Al$_{0.24}$Si$_{1.58}$, and between 1.9 K and 300 K for the non-magnetic La-analog. The data in the Ce compound exhibit a lambda type anomaly with a prominent peak at 9.3 K (taken as $T_C$), confirming the bulk magnetic ordering in CeCu$_{0.18}$Al$_{0.24}$Si$_{1.58}$ (Figure 4a). The jump in the heat capacity at $T_C$ (4.8 J/mol K) is less than 12.5 J/mol K, the mean field value for a doublet ground state with effective spin 1/2. This may imply either a substantial short range order in the paramagnetic state or partial compensation of the Ce 4$f$-spin by Kondo interaction. The jump in $C_P$ at $T_C$ further reduces with magnetic field (applied along [010]) and a broad hump-like feature, which shifts to higher temperatures with magnetic field, is observed in contrast to the sharp peak at the magnetic transition in zero field. These observations are in accordance with the ferromagnetic nature of ordering, possible short range order above $T_C$ and Kondo effect. The low temperature $C_p(T)/T$ vs. $T^2$ plot in the range 0.05-1 K is shown in the inset of Figure 4a, from where Sommerfeld coefficient, $\gamma$, is deduced to be 99.6 mJ/mol K$^2$ (in contrast to $\gamma$ = 3.2 mJ/mol K$^2$ for La-analog) by fitting the expression, $C_p/T = \gamma + \beta T^2$, below 1K. Note that $\gamma$ scales linearly with the electron effective mass $m^*$ in the Kondo problem. This moderately enhanced $\gamma$ and hence $m^*$ in CeCu$_{0.18}$Al$_{0.24}$Si$_{1.58}$ is presumably due to the hybridization between conduction electrons and Ce $f$-spins by the Kondo mechanism.

The 4$f$ electron contribution to the entropy as a function of temperature, $S_{4f}(T)$, calculated by using the thermodynamic relation, $S_{4f} = \int \frac{C_{4f}}{T} dT$, is shown in Figure4a by solid black trace. $C_{4f}$ was calculated by subtracting the heat capacity of the La analog from that of CeCu$_{0.18}$Al$_{0.24}$Si$_{1.58}$, assuming the phonon contributions to be identical in two compounds. The entropy released at $T_C$ is only 0.68 Rln2, while its value corresponding to full doublet ground state with effective spin 1/2 (*i.e.* Rln2) is attained at 36 K; far above $T_C$. The observed behavior of entropy is tentatively attributed to the combined effect of Kondo interaction and possible short range magnetic correlations persisting to temperatures above $T_C$.

$C_{4f}$ between 1.9 and 300 K in CeCu$_{0.18}$Al$_{0.24}$Si$_{1.58}$ is shown in Figure4b. It exhibits a broad Schottky anomaly centered at around 100 K, which arises due to the variation in the population of the crystal electric field levels with temperature governed by Boltzmann distribution function. $C_{4f}$ in the paramagnetic state was analyzed by a model presented in Ref. 3 by considering Kondo ($C_{Kondo}$) and Schottky ($C_{Schottky}$) contributions. From the least square fitting of the model presented in ref. 3 to $C_{4f}(T)$ data of CeCu$_{0.18}$Al$_{0.24}$Si$_{1.58}$ in the temperature range 11-300 K, we have determined that the excited crystal electric field (CEF) split doublets are located at 178 and 357 K, respectively, from the ground doublet and the Kondo temperature $T_K$ = 8.4 K. To confirm our estimate of the CEF level splittings inelastic neutron scattering is required. Our analysis shows that $T_K$ and $T_C$ are comparable in CeCu$_{0.18}$Al$_{0.24}$Si$_{1.58}$, making it a potential candidate for exhibiting a ferromagnetic quantum critical point. Towards that direction, we tried to measure the resistivity under pressure in a piston-cylinder type indenter cell, but due to technical difficulties mainly arising from the breaking of the lead contacts our attempts have not yet succeeded.

4.  **Conclusions**

We have grown a single crystal of CeCu$_{0.18}$Al$_{0.24}$Si$_{1.58}$ by flux method. From single crystal XRD analysis it is confirmed that this compound is a disordered derivative of the tetragonal $\alpha$-ThSi$_2$-type structure, with Al and Cu atoms partially replacing Si but with full occupation of the 8*e* Wyckoff site. The compound is ferromagnetic with a Curie temperature of 9.3 K. A hysteretic magnetization curve in the magnetically ordered state, reduced jump in the $C_p$ and magnetic entropy at $T_C$, negative temperature coefficient in electrical resistivity indicate that the present compound is a rare example of a Kondo lattice ordering ferromagnetically. An analysis to the $C_{4f}$ data located the CEF excited states at 178 K and 357 K, respectively, and Kondo temperature $T_K = 8.4$ K. We propose microscopic techniques like neutron scattering experiment to confirm these values. $CeCu_{0.18}Al_{0.24}Si_{1.58}$ is an interesting case to explore the quantum critical point by pressure or doping.

Table 1. Single crystal data and structure refinement details for CeCu$_{0.18}$Al$_{0.24}$Si$_{1.58}$ [$T$ = 293(2) K].

| **Compound** | CeCu$_{0.18}$Al$_{0.24}$Si$_{1.58}$ |
|---|---|
| Structural prototype | ThSi$_2$ |
| Pearson code | *tI*12 |
| Crystal system | Tetragonal |
| Space group | *I*4$_1$/*amd* (No. 141) |
| Lattice parameters [Å] | $a$ = 4.2160(10) |
| | $c$ = 14.147(2) |
| Unit cell volume [Å$^3$] | 251.46(9) |
| Unit formula per cell, Z | 4 |
| Absorption coefficient, μ (Mo K$\alpha$) [mm$^{-1}$] | 20.04 |
| Calculated density, ρ [g/cm$^3$] | 5.412 |
| Crystal size [μm] | 50 × 60 × 140 |
| Scan mode | ω-θ |
| *F*(000) | 354.1 |
| Theta range [°] | 2 ≤ θ ≤ 32 |
| Range in *h*, *k*, *l* | –6 ≤ *h* ≤ 6, –6 ≤ *k* ≤ 6, –21 ≤ *l* ≤ 0 |
| Measured reflections | 908 |
| Independent reflections | 137 |
| Absorption correction | Ψ-scans |
| Refinement method | Full-matrix least-squares on F$^2$ |
| Refined parameters | 8 |
| Extinction coefficient | 0.16(2) |
| Data restraints/parameters | 1 |
| *R*1, *wR*2 (F$_o^2$) over all data | 0.027, 0.048 |
| *R*1 for 114 refl. with I > 2 sigma(I) | 0.020 |
| Goodness of fit | 1.156 |
| Δρ$_{max}$, Δρ$_{min}$, e/Å$^3$ | + 1.79, – 2.02 |

Table 2. Standardized fractional atomic coordinates and displacement parameters of CeCu$_{0.18}$Al$_{0.24}$Si$_{1.58}$. The equivalent displacement parameter, $U_{eq.}$, is defined as one third of the trace of the orthogonalized $U_{ij}$ tensor. $U_{12} = U_{23} = U_{13} = 0$.

| Atom | Wyckoff site | Atomic coordinates | | | Occ. | $U_{eq}$ [Å$^2$] | $U_{11}$ [Å$^2$] | $U_{22}$ [Å$^2$] | $U_{33}$ [Å$^2$] |
|---|---|---|---|---|---|---|---|---|---|
| | | $x$ | $y$ | $z$ | | | | | |
| Ce | 4$a$ | 0 | 3/4 | 1/8 | 1 | 0.0073(4) | 0.0066(4) | 0.0066(4) | 0.0086(4) |
| Cu/Al/Si | 8$e$ | 0 | 1/4 | 0.2910(1) | 0.09/0.12/0.79* | 0.0093(4) | 0.0083(10) | 0.0109(10) | 0.0085(8) |

*values fixed according to averaged SEM-EDAX analyses. Final stoichiometry: CeCu$_{0.18}$Al$_{0.24}$Si$_{1.58}$.

Table 3. Bond distances in the CeCu$_{0.18}$Al$_{0.24}$Si$_{1.58}$ compound (M = Cu, Al, Si).

| Central atom | Ligands | d [Å] | d$_{obs}$/Σ$r_M$ |
|---|---|---|---|
| **Ce–** (CN = 12) | 4 M | 3.156 (2) | 1.005 |
| | 8 M | 3.209(1) | 1.022 |

| Central atom | Ligands | d [Å] | d$_{obs}$/Σ$r_M$ |
|---|---|---|---|
| **M–** (CN = 9) | 1 M | 2.376(4) | 0.903 |
| | 2 M | 2.407(2) | 0.915 |
| | 2 Ce | 3.156(2) | 1.005 |
| | 4 Ce | 3.209(1) | 1.022 |

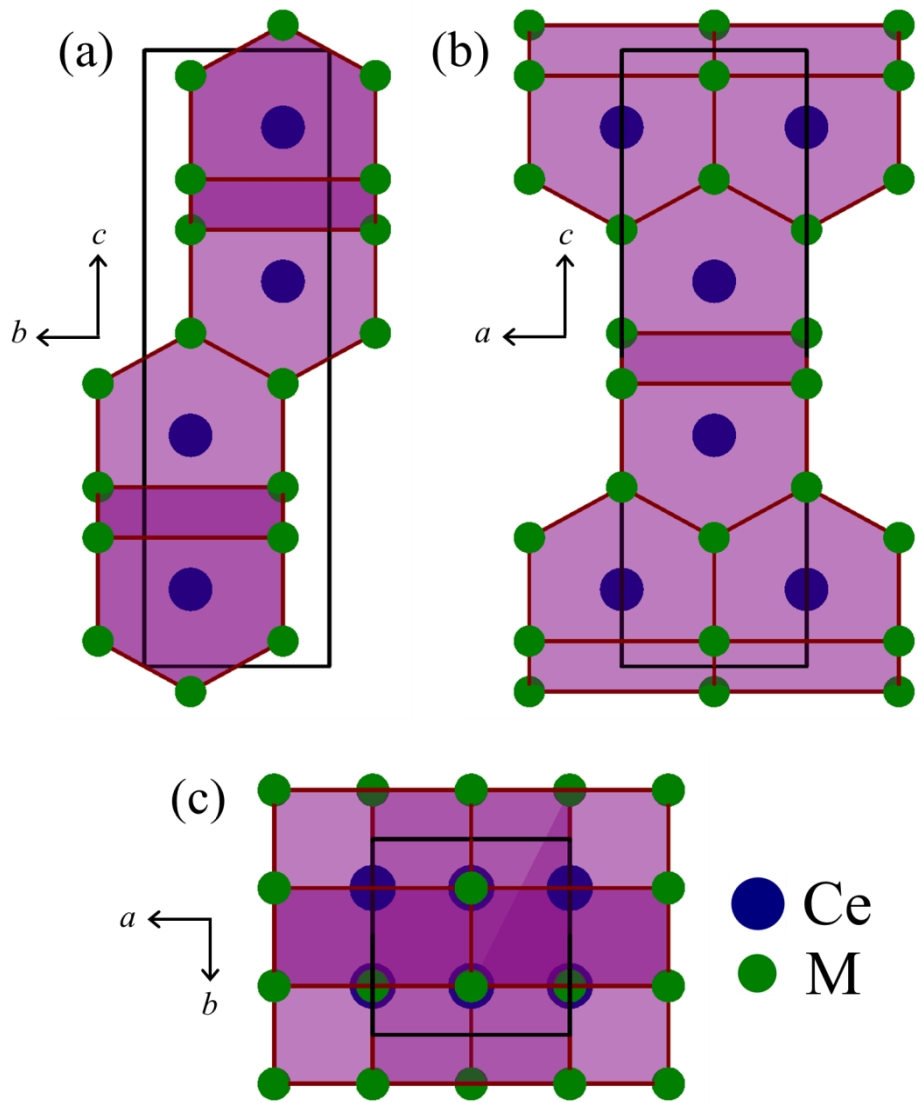

Figure 1. Projections of the unit cell of the CeCu$_{0.18}$Al$_{0.24}$Si$_{1.58}$ along the [100] (1a), [010] (1b) and [001] (1c) directions; the polyhedra around the Ce atoms are highlighted.

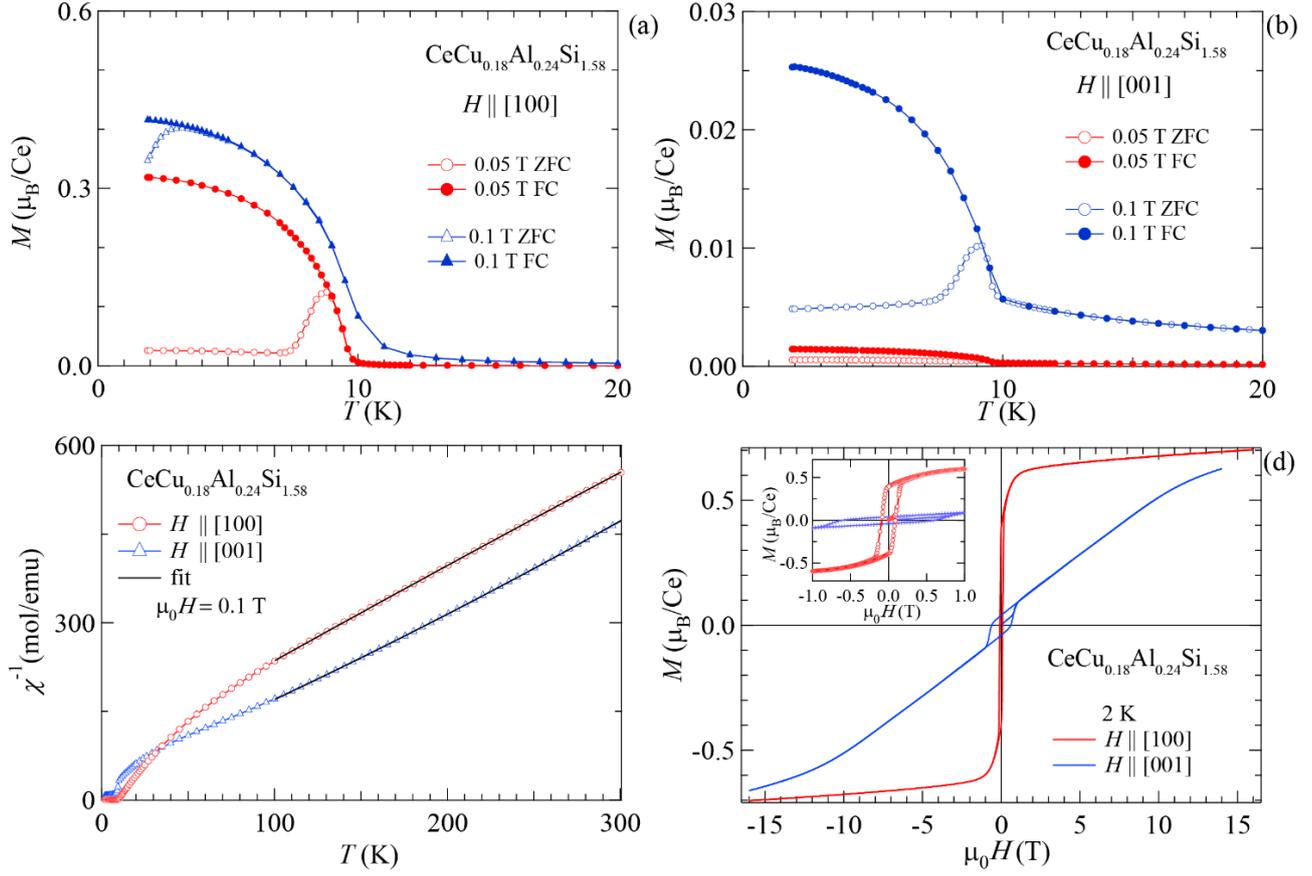

Figure 2. ZFC and FC magnetization *vs.* temperature, $M(T)$, measured in magnetic fields of 0.05 T and 0.1 T along the [100] (a) and [001] (b) directions, respectively; (c) inverse magnetic susceptibility, $\chi^{-1}$, as a function of temperature measured in a field of 0.1 T; (d) isothermal magnetization at 2 K measured along the principal crystallographic directions [100] and [001] (the inset shows the anisotropy in the hysteresis on an expanded scale).

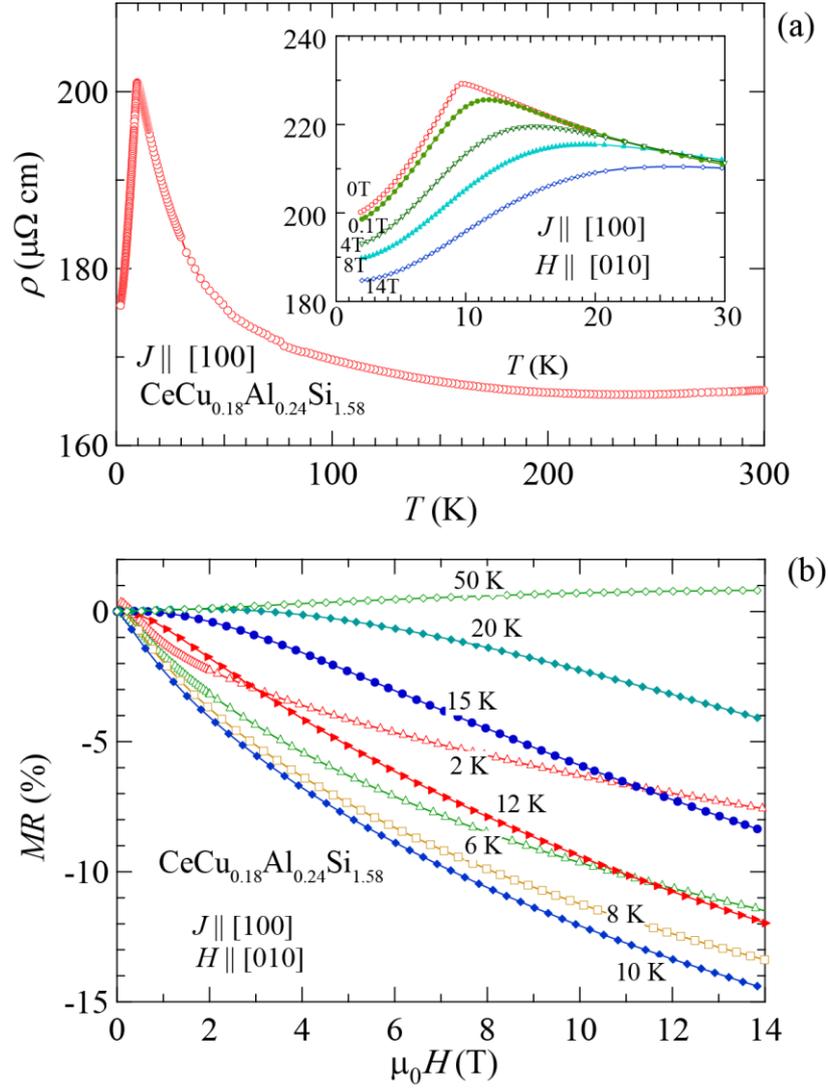

Figure 3 (a) Electrical resistivity as a function of temperature, $\rho$(T), of CeCu$_{0.18}$Al$_{0.24}$Si$_{1.58}$, for $J \parallel$ [100] in the main panel and at selected fields for $J \parallel$ [100] and $H \parallel$ [010] (in inset). (b) *MR*(*H*) at temperatures ranging from 2 to 50K for $J \parallel$ [100] and $H \parallel$ [010] of CeCu$_{0.18}$Al$_{0.24}$Si$_{1.58}$.

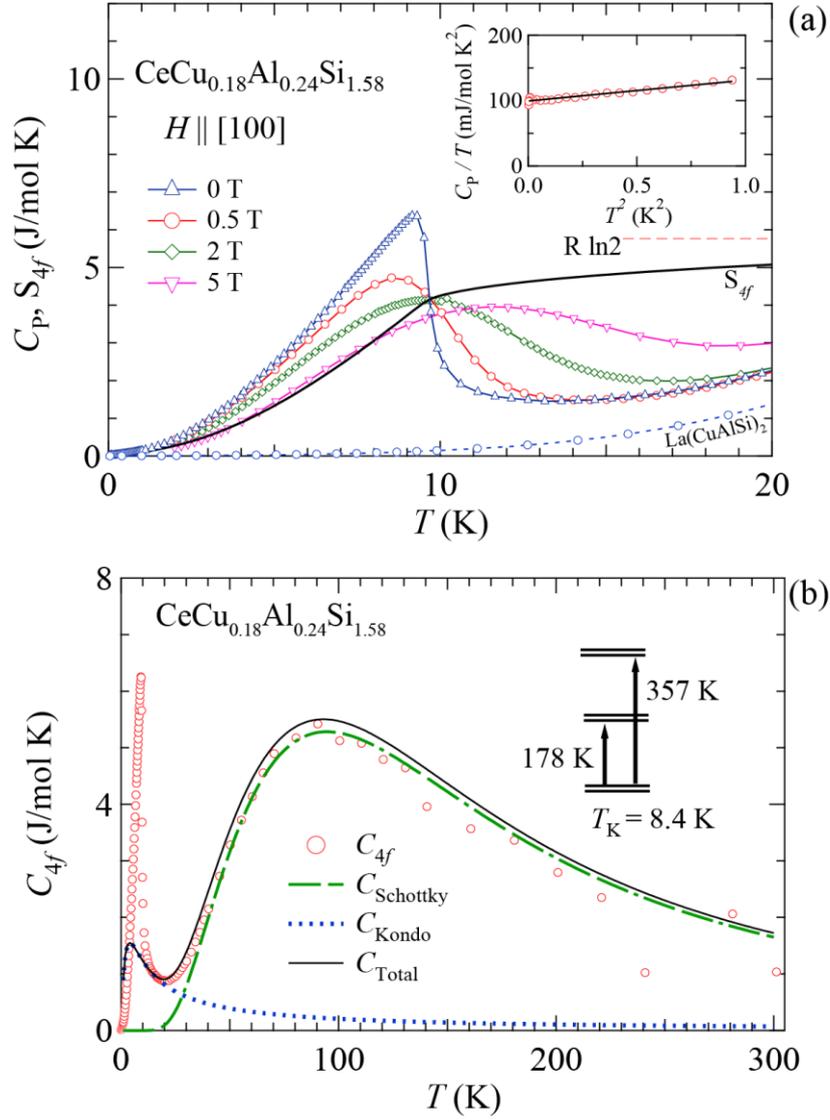

Figure 4.(a) Heat capacity, $C_P(T)$, of CeCu$_{0.18}$Al$_{0.24}$Si$_{1.58}$ measured at selected fields; the solid black line is the calculated entropy, $S_{4f}$ (the inset shows the low temperature $C_P$ data plotted as a function of $C_P/T$ vs. $T^2$ to deduce the electronic coefficient of heat capacity, $\gamma$). $C_P(T)$ of nonmagnetic analog LaCu$_{0.18}$Al$_{0.24}$Si$_{1.58}$ is also shown (b) $C_{4f}(T)$ depicting the Schottky anomaly and fitted model to deduce the gap in Ce$^{3+}$ CEF energy levels and $T_K$ (see text).